\documentclass[pra,aps,twocolumn,superscriptaddress,showpacs,eqsecnum,nofootinbib]{revtex4-1}

\usepackage{amsmath}
\usepackage{graphicx}
\usepackage{mathrsfs}
\usepackage{color}
\usepackage{hyperref}

\hypersetup{
	colorlinks = true,
	urlcolor   = blue,
	linkcolor  = red,
	citecolor  = blue
}

\begin{document}

\title{Quantum interface between light and a one-dimensional atomic system}

\author{V.A. Pivovarov}
\affiliation{Physics Department, St.-Petersburg Academic University, Khlopina 8, 194021 St.-Petersburg, Russia}
\affiliation{Center for Advanced Studies, Peter the Great St-Petersburg Polytechnic University, 195251, St.-Petersburg, Russia}
\author{A.S. Sheremet}
\affiliation{Russian Quantum Center, 121205 Skolkovo IC, Bolshoy Bulvar 30-1, Moscow Region, Russia}
\author{L.V. Gerasimov}
\affiliation{Quantum Technologies Center, M.V.~Lomonosov Moscow State University, Leninskiye Gory 1-35, 119991, Moscow, Russia}
\affiliation{Center for Advanced Studies, Peter the Great St-Petersburg Polytechnic University, 195251, St.-Petersburg, Russia}
\author{J. Laurat}
\affiliation{Laboratoire Kastler Brossel, Sorbonne Universit\'e, CNRS,
ENS-Universit\'e PSL, Coll\`ege de France, 4 place Jussieu, 75005 Paris, France}
\author{D.V. Kupriyanov}
\email{kupriyanov@quantum.msu.ru}
\affiliation{Center for Advanced Studies, Peter the Great St-Petersburg Polytechnic University, 195251, St.-Petersburg, Russia}
\affiliation{Quantum Technologies Center, M.V.~Lomonosov Moscow State University, Leninskiye Gory 1-35, 119991, Moscow, Russia}

\date{\today}
\begin{abstract}
\noindent We investigate optimal conditions for the quantum interface between a signal photon pulse and one-dimensional chain consisting of a varied number of atoms. The tested object is physically designed as an atomic array of tripod-type atoms confined with a nanoscale dielectric waveguide and experiencing its fundamental HE${}_{11}$ mode. The efficiency of interaction is mainly limited by achieved overlap and coupling of the waveguide evanescent field with the trapped atoms. We verify those physical conditions when the coupling within the main scattering channels would be sufficient for further providing various elementary interface protocols such as light storage, light-matter entanglement, preparation of a few photon states on demand, etc..
\end{abstract}


\maketitle

\section{Introduction}
\noindent Realization of noisy protected quantum communications with the concept of quantum repeater protocol \cite{Briegel1998,Kimble2008} needs in highly effective quantum memories as a key element for its technical implementation. Cold and ultracold atomic systems are considered as a convenient physical platform for providing controllable interchange by the quantum states between a signal light pulse and atomic register \cite{HamSorPol2010,Gisin2011}. Recent experiments have revealed that ensembles of alkali-metal atoms can demonstrate a long-lived and quite effective memory for qubit mapping onto the atomic ground state spin subsystem  \cite{Kuzmich2013,Laurat2015,Laurat2018,Wang2019}. With constructing stable configurations of atomic arrays that opens perspective for preparation of multiqubit quantum registers and for quantum simulations with avoiding decoherence in the multiqubit systems due to infinitely long natural relaxation time. In free space it is accomplished a near-deterministic loading up to hundred atoms in the microtraps with occupation a single atom per trap and arrangement of optical tweezers into lattices of arbitrary shape with holographic techniques \cite{Browaeys2014,Kaufman2015}. By integration with a nanofiber assisted technique it is allowed preparation of even more (up to thousands) atoms confined with the nanofiber waveguide and controlled by interaction with the evanescent field of the fundamental waveguide mode \cite{Balykin2005,Rauschenbeutel2010,Kimble2012}. The technique has shown a various of convenient options helpful for further implementation to the quantum interface between light and one-dimensional atomic lattices, such as anisotropic and dimensional controllable cooperative emission \cite{Rauschenbeutel2014} and strong cooperative light reflection \cite{CGCGSKL2016,SBKISMPA2016}.

From other hand, recent studies of coherent optical processes developing in cold and ultracold atomic systems have elaborated various of experimental methods for manipulation with collective atomic states at mesoscopic level and with quantum precision \cite{HamSorPol2010,KSH2017}. That motivated a number of supporting theoretical treatments towards complex description of the interaction processes developing between small atomic samples and a few photon states of light.

For the ensembles of cold atoms, existing in non-degenerate phase and approximated by collections of infinitely massive particles, the atom-field interaction is relevantly described in the dipole approximation. The microscopic description, beyond the self-consistent approach, can be developed if the fundamental solution of the Maxwell equation for a point-like dipole source can be constructed in analytical form and allows to evaluate the resolvent operator of the system Hamiltonian \cite{SMLK2012,SKGLK2015,Browaeys2016,PLGPCLK2018,Chang2018}. That cannot be so simply done in the case of atoms confined with a nanostructure and therefore has to be additionally motivated.

The important extensions of empirical description of the collective processes, primary based on the simplified assumption of two-level atoms, have indicated various manifestations of cooperative phenomena \cite{KorSherPetr2016,Javanainen2017,Chang2017} and quantum correlations \cite{Hammerer2018} in the one-dimensional atomic configurations. In two and three-level models with quantum description of the guided light, the action of the control field is normally considered via coupling with a signal level isolated from interaction with the waveguide \cite{Chang2017}. Nevertheless the testing of the EIT process, with precise polarization description of both  the guided field and control mode, has revealed its strong sensitivity to the mode structure and distribution of atoms \cite{Rauschenbeutel2015}. The microscopic analysis, beyond the self-consistent description of the atomic subsystem and presented for a closed and solvable vector model, would help in experimental verification of such important phenomena as quantum entanglement and cooperativity and guide us towards optimal conditions for the quantum interface.

In this paper we are aiming to elaborate the microscopic approach for an atomic system in the complete vector model for tripod-type atoms confined with a nanofiber waveguide, which we consider as a convenient
and realistic illustration of a one-dimensional scheme for any quantum interface protocol. In particular the abilities of the quantum memory protocol will be examined by numerical simulations based on the formalism of the quantum scattering theory. We consider the standard configuration for the scattering of a single photon wave packet from an atomic array in the presence of the control mode driving the empty transition.  The atoms are treated as infinitely massive immobile particles distributed along the waveguide in either ordered or disordered configuration and separated by a distance either exactly or approximately fulfilling the conditions of the Bragg coherent scattering. As it can be expected, such a deviation in distributions would lead to dramatic changes in manifestation of the cooperative effects.

We can precede our discussion by the following suggestive arguments with pointing the important difference with the similar scattering process developing in free space. For a non-saturating light pulse its propagation through an atomic sample in diffusion regime obeys the Beer-Lambert-Bouguer law, which is valid for each segment of its ballistic passages. The spectrum of incoherent losses near the resonance is scaled as $\gamma\sqrt{b_0}$, where $\gamma$ is a single atom spontaneous decay rate and $b_0$ is optical depth of the sample at the resonance point. For a one-dimensional system confined with a waveguide we can apply similar estimate for the incoherent losses with substituting the optical depth by the total number of atoms in the chain $N$, and with expanding the spontaneous decay only on external modes with a rate $\gamma^{\mathrm{ext}}$, so the spectrum is scaled as $\gamma^{\mathrm{ext}}\sqrt{N}$. From other hand, the rate of cooperative emission from a one-dimensional Dicke-type atomic system, consisting of $N$ atoms, is scaled as $\gamma^{\mathrm{wg}}N$ where $\gamma^{\mathrm{wg}}$ is emission rate per atom into the waveguide mode. Typically $\gamma^{\mathrm{ext}}\sim\gamma\gg\gamma^{\mathrm{wg}}$ (a small factor of Purcell enhancement) such that for a single atom the light is mostly scattered into the external modes. But the balance could be changed once the atomic chain was prepared with a sufficiently large number of atoms if $N>(\gamma/\gamma^{\mathrm{wg}})^2$. Then one can expect a highly effective interaction between the signal pulse and atomic array by communication via the waveguide and with minimizing the losses. We are aiming to demonstrate such a scenario in the presence of the control field by a round of microscopic \textit{ab-initio} numerical simulations presented for a collection of tripod-type atoms confined with a nanoscale dielectric waveguide.

The paper is organized in two parts. In Section \ref{Section_II} we overview our calculation approach and in Appendix \ref{Appendix_A} we show how the photon Green's function can be corrected for light propagation near a nanofiber structure. In Section \ref{Section_III} we present the results of our numerical simulations and discuss how the memory effect depends on the scattering geometry and on the number of atoms. We summarize our main results in Conclusion.

\section{Theory}\label{Section_II}
\noindent In this section we generalize the theoretical approach previously developed in \cite{PLGPCLK2018} towards (i) examination of the scattering process in its dependence on the number of atoms and their distribution in the chain and (ii) introducing the control field. For detail definitions, notations and the links with the standard formalism of the quantum scattering theory, which we will follow here, we readdress the reader to \cite{SMLK2012,SKGLK2015,PLGPCLK2018}.

\subsection{The scattering matrix and resolvent operator}
\noindent In accordance with general principles of the scattering theory the dynamics of a single photon wave-packet interacting with an atomic sample can be described in formalism of the scattering $S$-matrix that transforms the system states from infinite past $|\psi\rangle_{\mathrm{in}}$ to infinite future $|\psi\rangle_{\mathrm{out}}$ as a result of the interaction process \cite{GoldbergerWatson64}. In the interaction representation, the corresponding asymptotic transformation is given by
\begin{equation}
|\psi\rangle_{\mathrm{out}}=\mathrm{e}^{\frac{i}{2\hbar}H_0\tau}\mathrm{e}^{-\frac{i}{\hbar}H\,\tau}\mathrm{e}^{\frac{i}{2\hbar}H_0\tau}|\psi\rangle_{\mathrm{in}}\equiv%
\hat{S}|\psi\rangle_{\mathrm{in}},
\label{2.1}
\end{equation}
where $\tau\to+\infty$, $H$ is the system Hamiltonian and $H_0$ is its non-interacting part. The operator $\hat{S}$ can be represented as a matrix in a decoupled basis of two interacting subsystems, which we specify as $|\phi_i\rangle$ for the initial and $|\phi_{i'}\rangle$ for the final system states. As shown in \cite{PLGPCLK2018} for one-dimensional system the representative matrix elements of the $S$-matrix are given by
\begin{eqnarray}
S_{i'i}&=&\delta_{i'i}-i\,\frac{{\cal L}}{\hbar v_g}\,T_{i'i}(E_i+i0),%
\label{2.2}
\end{eqnarray}
where the initial and final states $i\equiv g,s$ and $i'\equiv g',s'$ specify the scattering within the waveguide modes $s\to s'$ and with changing of the internal collective spin state in atomic system $g\to g'$. The $S$-matrix is parameterized by a length of quantization segment ${\cal L}$ (defining the longitudinal mode structure, see below) and the group velocity $v_g$ assists the free propagation of the incoming and outgoing wave packets along the fiber before and after interaction. The scattering dynamics is described by the $T$-matrix, contributed to the second term in (\ref{2.2}), which can be expanded in the perturbation theory series and than can be calculated by the Feynman diagram method.

In the case of a near-resonance scattering the $T$-matrix elements are given by
\begin{eqnarray}
\lefteqn{T_{g's',g\,s}(E)=2\pi\hbar\sqrt{\omega_{s'}\omega_s}}
\nonumber\\%
&&\times\sum_{b,a=1}^{N}\;\sum_{n',n}%
\left(\mathbf{d}\!\cdot\!\mathbf{D}^{(s')}(\mathbf{r}_b)\right)_{n'm'_b}^{*}\left(\mathbf{d}\!\cdot\!\mathbf{E}^{(s)}(\mathbf{r}_a)\right)_{nm_a}%
\nonumber\\%
&&\times\langle\ldots m'_{b-1},n',m'_{b+1}\ldots |\tilde{\hat{R}}(E)%
|\ldots m_{a-1},n,m_{a+1}\ldots \rangle,%
\nonumber\\%
&&\label{2.3}%
\end{eqnarray}
where $\omega_s$ and $\omega_{s'}$ are the frequencies of the incident and scattered photons respectively. \footnote{We will call such a quasi-particle as photon, but strictly saying it is a polariton wave propagating through a dielectric waveguide, but we shall reserve the term "polariton" for the superposition of this wave with a single atom excitation, see below.}  The incident mode $s$ contributes by its electric field profile $\mathbf{E}^{(s)}(\mathbf{r})$ and the outgoing mode $s'$ contributes by the profile of its displacement field $\mathbf{D}^{(s')}(\mathbf{r})$. The incident photon is annihilated and the outgoing photon is created at location points $\mathbf{r}_{a}$ and $\mathbf{r}_{b}$ of arbitrary atoms of the atomic chain. The transition amplitude is intrinsically determined by the matrix element of the resolvent operator of the system Hamiltonian projected onto a collective atomic state with a single optical excitation
\begin{equation}
\tilde{\hat{R}}(E)=\hat{P}\,\hat{R}(E)\,\hat{P}\equiv \hat{P}\frac{1}{E-\hat{H}}\hat{P}\,.%
\label{2.4}%
\end{equation}
The projector $\hat{P}$ is given by
\begin{eqnarray}
\lefteqn{\hspace{-0.8cm}\hat{P}=\sum_{a=1}^{N}\;\sum_{\{m_j\},j\neq a}\;\sum_{n}%
|m_1,\ldots,m_{a-1},n,m_{a+1},\ldots m_N\rangle}%
\nonumber\\%
&&\hspace{-0.5cm}\langle m_1,\ldots,m_{a-1},n,m_{a+1},\ldots m_N|\times|0\rangle\langle 0|_{\mathrm{Field}}%
\label{2.5}%
\end{eqnarray}
and selects in the atomic Hilbert subspace the entire set of the states where any $j$-th of $N-1$ atoms populates a Zeeman sublevel $|m_j\rangle$ in its ground state and one specific $a$-th atom (with $a$ running from $1$ to $N$ and $j\neq a$) populates a Zeeman sublevel $|n\rangle$ of its excited state. The field subspace is projected onto its vacuum state and operator $\tilde{\hat{R}}(E)$ can be further considered as a matrix operator acting only in the atomic subspace. In the representation of the $T$-matrix by the expansion (\ref{2.3}) the selected specific product of matrix elements runs all the possibilities when the incoming photon is annihilated on any $a$-th atom and the outgoing photon is created on any $b$-th atom of ensemble, including the possible coincidence $a=b$. The initial atomic state is given by $|g\rangle\equiv|m_1,\ldots,m_N\rangle$ and the final atomic state by $|g'\rangle\equiv|m'_1,\ldots,m'_N\rangle$, where atoms can populate all the accessible internal states. The normalization length ${\cal L}$ cancels out when substituting (\ref{2.3}) into (\ref{2.2}), so the elements of $S$-matrix (\ref{2.2}) give us the set of the quantum probability amplitudes for observation of the system in particular final states in the quasi-one-dimensional scattering process.

For the system consisting of many atoms with degenerate ground state there is an exponentially rising up number of the scattering channels. Hopefully for most of the problems associated with quantum interface, such as quantum swapping, memories, entanglement etc., the elastic scattering channel and the channels with minimal number of Raman transitions are mostly important. This significantly simplifies the problem with constructing the resolvent operator (\ref{2.4}) with letting us operate in a representative part of the Hilbert subspace, see comments below in section \ref{II.D}. The critical step in solving many particle problem is to build up the Green's function of the electric field, which is identified as the fundamental solution of the Maxwell equations for a point-like dipole source located near a nanostructure.

\subsection{The electric field Green's function}
\noindent As proven in statistical physics, see \cite{LfPtIX}, the causal-type electric field Green's function considered in a spatial region nearby a macroscopic object can be expressed by the retarded-type fundamental solution of the macroscopic Maxwell equations
\begin{eqnarray}
D^{(E)}_{\mu\nu}(\mathbf{r},\mathbf{r}';\omega)\!\!&=&\!\!-i\!\int^{\infty}_{-\infty}\!\!d\tau\,%
\mathrm{e}^{i\omega\tau}\!\left.\langle T E_{\mu}(\mathbf{r},t)\,E_{\nu}(\mathbf{r}',t')\rangle\right|_{\tau=t-t'}%
\nonumber\\%
\!&=&\!\frac{\omega^2}{c^2}D^{(R)}_{\mu\nu}(\mathbf{r},\mathbf{r}';|\omega|),%
\label{2.6}
\end{eqnarray}
where the $D^{(R)}$ function is the fundamental solution for a point-like charge distribution and the $D^{(E)}$ function reproduces the electric field emitted by a point-like dipole source. It is crucially for the paradigm of statistical physics that the integrand in the first line of (\ref{2.6}) is associated with expectation value of a microscopically defined time-ordered product (pointed by $T$-symbol) of the electric field operators in the Heisenberg representation. In the considered case the $D^{(E)}$ function is given by the sum of two contributions
\begin{equation}
D^{(E)}_{\mu\nu}(\mathbf{r},\mathbf{r}';\omega)=D^{(\mathrm{wg})}_{\mu\nu}(\mathbf{r},\mathbf{r}';\omega)+D^{(\mathrm{ext})}_{\mu\nu}(\mathbf{r},\mathbf{r}';\omega).
\label{2.7}
\end{equation}
The first term is contribution of the waveguide modes, which is given by
\begin{equation}
D^{(\mathrm{wg})}_{\mu\nu}(\mathbf{r},\mathbf{r}';\omega)=\sum_{s}\frac{4\pi\hbar\omega^2}{\omega^2-\omega_{s}^{2}+i0}E^{(s)}_{\mu}(\mathbf{r})\,E^{(s)\ast}_{\nu}(\mathbf{r}').%
\label{2.8}
\end{equation}
Here we have introduced the $\mu$-th and $\nu$-th Cartesian components of the vector mode functions $\mathbf{E}^{(s)}(\mathbf{r})$. These functions should be found via respective solution of the homogeneous Maxwell equations and fulfill the following normalization conditions
\begin{eqnarray}
\lefteqn{\int\! d^3r\,\epsilon(\mathbf{r})\, \mathbf{E}^{(s')\ast}(\mathbf{r})\cdot\mathbf{E}^{(s)}(\mathbf{r})}
\nonumber\\%
&&\equiv\int\! d^3r\,\mathbf{D}^{(s')\ast}(\mathbf{r})\cdot\mathbf{E}^{(s)}(\mathbf{r})=\delta_{s's},%
\label{2.9}
\end{eqnarray}
where $\epsilon(\mathbf{r})=1+4\pi\chi(\mathbf{r})$ is the dielectric permittivity and $\mathbf{D}^{(s')}(\mathbf{r})=\epsilon(\mathbf{r})\, \mathbf{E}^{(s')}(\mathbf{r})$ is the displacement field of the $s'$-mode. Each mode is parameterized by entire mode index $s=\sigma,k$ where $\sigma=\pm 1$ is its azimuthal quantum number and $k$ is its longitudinal wave number. The integral over the longitudinal $z$-variable is bounded by the quantization segment ${\cal L}\to\infty$ and implies periodic boundary conditions and quasi-discrete spectrum of $k$.

The term $D^{(\mathrm{ext})}$ in Eq.~(\ref{2.7}) is the contribution of light emission into the external modes. For a point-like dipole source, separated from the fiber surface by a distance comparable with the fiber diameter of sub-wavelength scale, this term can be approximated  by the vacuum Green's function  $D^{(0)}$ slightly distorted by the presence of the waveguide. In Appendix \ref{Appendix_A} we show how this term can be constructed in analytical form accordingly the basic principles of the scattering theory.

\subsection{Signal light, mode structure, control field}

\noindent We consider the signal light as a single-photon pulse propagating through the waveguide and expanded in its fundamental mode. There are two degenerate modes specified by the index $s=\sigma, k$, distinguished by their azimuthal numbers $\sigma=\pm 1$. In cylindric coordinates $\rho,\phi,z$ the mode components (the positive frequency components of the electric field) can be factorized as
\begin{equation}
E^{(s)}_{q}(\mathbf{r})=E^{(\sigma k)}_{q}(\rho)\,\frac{1}{\sqrt{2\pi {\cal L}}}\mathrm{e}^{i\sigma\phi}\,\mathrm{e}^{ik z},%
\label{2.10}%
\end{equation}
where the vector projections are referred with the local basis, which directors are enumerated by $q=\rho,\phi,z$. The two azimuthal modes can be expressed by three basic functions
\begin{eqnarray}
E_{\rho}^{(\pm 1 k)}(\rho)&=&E_{\rho}(\rho)%
\nonumber\\%
E_{\phi}^{(\pm 1 k)}(\rho)&=&\pm E_{\phi}(\rho)%
\nonumber\\%
E_{z}^{(\pm 1 k)}(\rho)&=&E_{z}(\rho),%
\label{2.11}%
\end{eqnarray}
which can be explicitly constructed as a combination of the Bessel functions by solution of the homogeneous Maxwell equations, see \cite{Marcuse82,PLGPCLK2018}.

It is more convenient to expand the modes in the Cartesian basis, directly connected with definitions of the atomic eigenstates. Then the electric field components are given by
\begin{eqnarray}
E_{x}^{(\pm 1 k)}(\rho,\phi)&=&\frac{E_{\rho}(\rho)-iE_{\phi}(\rho)}{2\sqrt{2\pi}}+\frac{E_{\rho}(\rho)+iE_{\phi}(\rho)}{2\sqrt{2\pi}}\,\mathrm{e}^{\pm 2i\phi}%
\nonumber\\%
E_{y}^{(\pm 1 k)}(\rho,\phi)&=&\pm\frac{iE_{\rho}(\rho)+E_{\phi}(\rho)}{2\sqrt{2\pi}}\mp\frac{iE_{\rho}(\rho)-E_{\phi}(\rho)}{2\sqrt{2\pi}}\,\mathrm{e}^{\pm 2i\phi}%
\nonumber\\%
E_{z}^{(\pm 1 k)}(\rho,\phi)&=&\frac{E_{z}(\rho)}{\sqrt{2\pi}}\,\mathrm{e}^{\pm i\phi}.%
\label{2.12}%
\end{eqnarray}
We will further assume the field distribution in the HE${}_{11}$-mode as mostly concentrated outside the waveguide with quite extended evanescent part. The mode is constructed as a superposition of the dominating transverse contributions, expressed by the first terms in the first and second lines of Eq.~(\ref{2.12}), with additional components possessing an orbital angular momentum. The orbital angular momentum induces precession to the Poynting vector of the propagating wave, which makes important difference for the waveguide mode (\ref{2.12}) with a plane wave propagating in free space.

\begin{figure}[tp]
{$\scalebox{0.5}{\includegraphics*{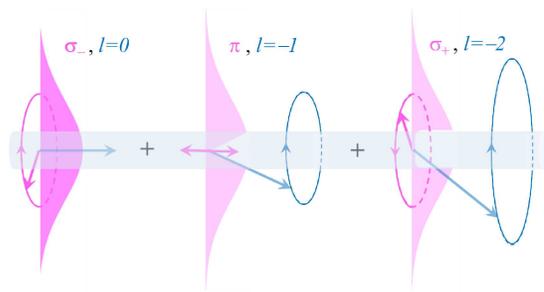}}$}
\caption{(color online).  Visualization of the fundamental HE${}_{11}$-mode with the azimutal number $\sigma=-1$ (total angular momentum of the mode).  The mode is superposed in three terms having different and mutually orthogonal polarizations $\sigma_{-}$, $\pi$ and $\sigma_{+}$, which gives the set of the internal (spin) angular momenta of the mode. The contributing terms have orbital angular momenta $l=0$, $l=-1$ and $l=-2$ respectively and provide the conservation of the total angular momentum. Only the dominant $\sigma_{-},l=0$ term survives the limit of an infinitely thin dielectric fiber and transforms to a plane wave propagating in free space.}
\label{fig1}%
\end{figure}

The specifics of the fundamental waveguide mode is clarified in the diagram of Fig.~\ref{fig1} in example of the $\sigma=-1$ azimuthal component and with regard to the optical transitions of a tripod-type atom. In the Cartesian frame associated with the atomic transitions there are three orthogonal polarization components conventionally named as $\sigma_{-}$ (left-handed), $\pi$ (longitudinal) and $\sigma_{+}$ (right-handed). In accordance with (\ref{2.12}) the diagram shows that the $\sigma=-1$ mode is superposed in three terms having the particular field polarizations and orbital angular momenta $l=0,-1,-2$. The internal (spin) angular momentum, given by rotation of the electric field vector, and the orbital angular momentum contribute to the total angular momentum equal to $\sigma=-1$ for all the three terms. For an infinitely thin dielectric fiber the mode structure is expected to approach the plane wave propagating in free space with surviving only the main contribution $\sigma_{-}, l=0$.

\begin{figure}[tp]
{$\scalebox{0.45}{\includegraphics*{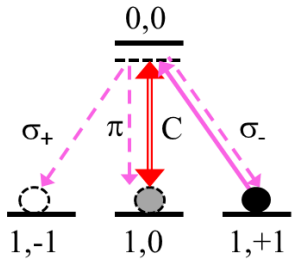}}$}
\caption{(color online). The transition diagram for the light storage protocol of a signal light pulse (magenta solid arrow) propagating in $\sigma=-1$ azimuthal mode and converted onto the atomic spin subsystem via the pulse scattering into the control mode (double red arrow). In this example the atoms are spin oriented along $z$-axis and the control beam $C$ is linearly polarized along $z$-axis and directed perpendicularly to the fiber. The dashed arrows indicate the spontaneous losses.}
\label{fig2}%
\end{figure}

The considered tripod-type atom has three Zeeman sublevels $F_0=1,M_0=0,\pm 1$ in its ground state and one level in its upper state $F=0,M=0$, as shown in Fig.~\ref{fig2}. Here we specify the atomic states by their total spin angular momenta $F_0,F$ and their projections $M_0,M$.  Such a closed energy structure exists in the hyperfine manifold of ${}^{87}$Rb. As follows from (\ref{2.12}) and, in example of $\sigma=-1$ mode, visualized by the diagram of Fig.~\ref{fig1} each of the azimuthal modes can actually excite any optical transition of the tripod-type atom, but with essentially different oscillator strengths. The $\sigma=+1$ and $\sigma=-1$ mode would most effectively drive the atom respectively at $\sigma_{+}$ and $\sigma_{-}$ transitions linked with the dominating circularly polarized transverse waves, expressed by the first terms in the first and second lines of (\ref{2.12}). If the atom is spin oriented along the waveguide and occupying the $M_0=+1$ state then the $\sigma=-1$ mode can drive only $\sigma_{-}$ excitation channel as shown in the transition diagram of Fig.~\ref{fig2}.

Nevertheless in the entire dynamics the incoincidence of the azimuthal number with the transition type makes a difference for the light scattering from an atom placed near waveguide than in free space. As a consequence of (\ref{2.12}) the Rayleigh scattering in either forward or backward directions, i.e. an event of the photon scattering within the waveguide modes when the atom stays in the same initial spin state, can happen with changing the azimuthal mode of the scattered photon. Furthermore, the Raman scattering, when the atom makes transition to another spin state, can happen with preserving the azimuthal mode of the scattered photon. In the case of spin-oriented atomic chain and for typical parameters of the nanofiber structures such options have low probability but not negligible and can be taken into account in the first order correction in the numerical evaluation of the resolvent operator (\ref{2.4}).

Any quantum state of the signal light can be mapped onto the atomic spin subsystem  with aid of a control field. The Raman gates providing the state conversion are shown in Fig.~\ref{fig2}. The atomic array is prepared as spin oriented along the $z$-direction and can effectively interact the signal light propagating in $\sigma=-1$ azimuthal mode via $\sigma_{-}$ transition. Once the control field is turned-on the state of the signal pulse can be converted onto the spin coherence in the atomic subsystem. The control field can exist as external plane wave, negligibly overlapping with the waveguide area, and is assumed as linearly polarized along $z$-axis and directed perpendicularly to the fiber, as it is shown in Fig.~\ref{fig3} and Fig.~\ref{fig7} below. We generally accept two options when the carrier frequency of the signal pulse is tuned either at the Autler-Townes (AT) absorption resonance, created by the control field, or alternatively at the electromagnetically induced transparency (EIT) point associated with this field.

The process can be introduced in the structure of the polariton propagator (resolvent operator) by adding to its self-energy part the following Feynman diagram\footnote{The diagram representation of the generalized Dyson equation for a collective propagator of the multiatomic system, where this graph should be inserted, is given in \cite{SMLK2012,PLGPCLK2018}}
\begin{equation}
\raisebox{-0.2 cm}{\scalebox{0.4}{\includegraphics*{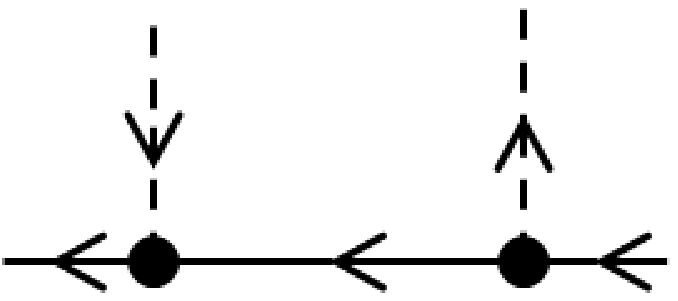}}}\ \Rightarrow\ -\frac{\hbar^2|\Omega_R|^2}{4\left(E-E_0-\hbar\omega_c+i0\right)}.%
\label{2.13}
\end{equation}
Here the internal solid line is the vacuum atomic propagator in the ground state, which is undisturbed by the presence of the waveguide. The inward and outward dashed arrowed lines are respectively the complex and complex conjugated amplitudes of the control field with frequency $\omega_c$ and with the Rabi frequency $\Omega_R$. The short incoming and outgoing arrows indicate the link with other terms of the diagram expansion for the dressed propagator of the polariton wave. Here $E$ is the energy argument in its Fourier representation and $E_0$ is the energy of the atomic ground state.

The key requirement justifying the validity of this assumption is that the diagram (\ref{2.13}) should not interfere with other terms contributing to the self-energy part of the polariton propagator, see \cite{PLGPCLK2018}. Mathematically, the cooperative scattering process from an atomic chain can be visualized by series of its perturbation theory expansion, which couple different pairs of atoms via virtual photons transporting the optical excitation through the chain. For each diagram the time delay between any two points, coupled by the Green's function of the photon, has an order of $L/c$ where $L$ is the chain length and $c$ is speed of light. In order to ignore the overlap between (\ref{2.13}) with other diagrams, contributing into the self-energy part, the control field amplitude should be sufficiently small such that $\Omega_{R}L/c\ll 1$. For the Rabi frequencies of the control field, typically comparable with a few $\gamma$, this inequality is surely fulfilled.

\subsection{Other approximations}\label{II.D}

\noindent The hardest point in evaluation of the resolvent operator in general case is in exponentially expanding dimension of the Hilbert subspace where the numerical simulations should be processed, which (with keeping only single optical excitation) is scaled as $Nd^{N-1}$, where $N$ is a number of atoms and $d=3$ is the degeneracy of the ground state. We can get around the problem if we keep only the representative domain of the full Hilbert space and indeed there are physical arguments that certain truncation to a subspace of less dimension can be done. Imagine that we have only one atom in the chain repopulated from the main state $F_0=1,M_0=1$ to any other Zeeman sublevel. Such an atom would experience only extremely weak interaction with a signal pulse propagating in $\sigma=-1$ azimuthal waveguide mode and for the considered calculation parameters the respective transmission losses can be numerically estimated by probability $\delta T\lesssim 0.005$. We use this bound to show how the full Hilbert space can be truncated to the representative subspace.

In the entire scattering process the atom can depopulate the $F_0=1,M_0=1$ state in result of either spontaneous Raman-type transition or via stimulated scattering into the control mode. The former process contributes into the self-energy part and into the resolvent operator (\ref{2.4}) by a sequence of virtual excitation transfers in its Feynman diagram expansion. For such virtual transitions and for an atomic chain consisting of about hundred atoms we keep the atomic repopulation to the states $F_0=1,M_0=0,-1$ only once. This can be justified by the above estimate of the individual transition probability and lets us describe the scattering process in the truncated Hilbert subspace with dimension $N+N(d-1)C_{N-1}^1=N+N(N-1)(d-1)$. Furthermore for the second process we are strongly constrained by our basic concept that the interaction of the control field with the atomic chain is allowed only within coherent mechanism described by the diagram (\ref{2.13}). Other dressing diagrams associated with either independent (non-cooperative) scattering or four-wave mixing of the control field interacting with the repopulated atoms turn the problem beyond this concept and should be ignored.  The validity of this can be justified by general weakness of the non-cooperative interaction of the control mode with any atom randomly repopulated onto $F_0=1,M_0=0$ state and would give only negligible correction to the calculated parameters for the main scattering channel. The diagram (\ref{2.13}) keeps the main cooperative part of the coherent interaction and leaves all the constructed matrix elements of the resolvent operator within the truncated Hilbert subspace.

\section{Results}\label{Section_III}

\noindent Below we present the results of our numerical simulations for two complementary geometries. In the first round of our simulations we follow the experimental design of \cite{CGCGSKL2016} and consider the scattering of the signal light incident on the atomic array from one direction. In the second round we split the signal pulse in a superposition of two counter-propagating waves and calculate the system response in a symmetric excitation geometry. For both the configurations we follow the signatures of the cooperativity in the system response and in enhancement of the Purcell effect.

\subsection{Single entry geometry}
\begin{figure}[tp]
{$\scalebox{0.6}{\includegraphics*{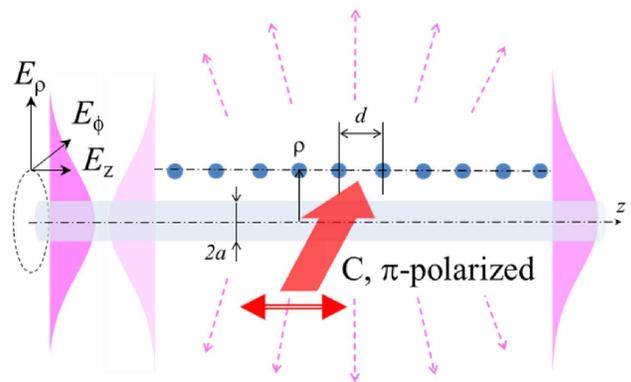}}$}
\caption{(color online). The geometry of signal light scattering from an array of atoms, separated by distance $d$ and confined with a nonofiber waveguide of radius $a$. The signal light impinges the target system in $\sigma=-1$ azumuthal mode and the control field is $\pi$-polarized and directed perpendicular to the fiber, see the transition scheme in Fig.~\ref{fig2}. The atoms, separated from the nonofiber surface by a distance $\rho-a$, experince the interaction with both the evanascent field of the HE$_{11}$-mode and with the control field.}
\label{fig3}%
\end{figure}

\noindent Let us consider the scattering process in geometry shown in Fig.~\ref{fig3}. The signal light in $\sigma=-1$ azimuthal component of the HE${}_{11}$-mode impinges on an atomic array, confined with a nanofiber dielectric waveguide, and splits by such a complex system into the transmitted and reflected fragments. A part of the light is lost because of the incoherent scattering into the external modes. The atoms are spin-oriented along the waveguide such that the interaction with the control field is triggered once the signal pulse is arrived, see the diagram of Fig.~\ref{fig2}.

The scattering process is described by coefficients of transmission $\mathscr{T}$, reflection $\mathscr{R}$ and losses $\mathscr{L}$, which are defined as
\begin{eqnarray}
\mathscr{T}&=&\mathscr{T}(\omega)=\sum_{i',k'>0}\left|S_{i'i}\right|^2%
\nonumber\\%
\mathscr{R}&=&\mathscr{R}(\omega)=\sum_{i',k'<0}\left|S_{i'i}\right|^2%
\nonumber\\%
\mathscr{L}&=&\mathscr{L}(\omega)=1-\mathscr{R}(\omega)-\mathscr{T}(\omega)
\label{3.1}%
\end{eqnarray}
and considered as function of the signal mode frequency $\omega\equiv\omega_s$. Here  $i=\{\sigma=-1,k;M_0=+1 (\mathrm{all\ atoms})\}$ and the final state can be any of $i'=\{\sigma',k';{\{M^{(a)\prime}_0\}}_{a=1}^N\}$ where each of the $N$ atoms can be redistributed onto arbitrary Zeeman sublevel with $M^{(a)\prime}_0=0,\pm 1$. However, as explained in the preceding section, for leading correction to the main scattering channels it is sufficient to keep only one event of the spontaneous Raman transition so we implies $M^{(a)\prime}_0\neq 1$ for only one atom in the chain.

\begin{figure}[tp]
{$\scalebox{0.5}{\includegraphics*{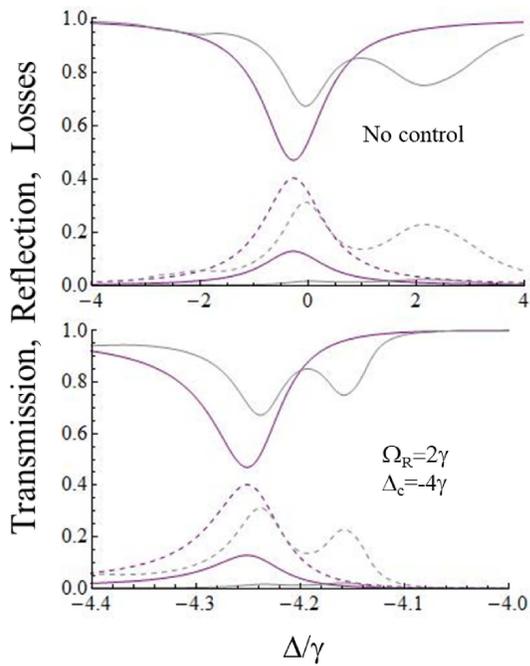}}$}
\caption{(color online) Spectra of transmission $\mathscr{T}=\mathscr{T}(\omega)$, reflection $\mathscr{R}=\mathscr{R}(\omega)$ and losses $\mathscr{L}=\mathscr{L}(\omega)$ for light scattered from an atomic chain consisting of ten atoms without (upper plot) and in the pesence (lower plot) of the control field. The spectra are plotted as function of detuning of the signal mode from the atomic resonance $\Delta=\omega-\omega_0$, see Fig.~\ref{fig2}, and presented for the ordered (magenta) and disordered (gray) atomic configurations with interatomic separation $d\sim\lambda^{\mathrm{wg}}/2$, see Fig.~\ref{fig3}. The solid curves correspond to the transmission and reflection and the dashed curves indicate the losses. The control mode $\omega_c$, having the Rabi frequency $\Omega_R=2\gamma$ and detuned from the atomic resonance by $\Delta_c=\omega_c-\omega_0=-4\gamma$, creates an artificial Autler-Townes resonance structure, which shows behavior similar to the scattering spectra near the fundamental matter-state atomic resonance.}
\label{fig4}%
\end{figure}

In Fig.~\ref{fig4} we show the spectra of transmission and reflection for an array consisting of ten atoms separated from the fiber surface by a distance $\rho-a=0.5a$, where $a$ is the fiber radius. In our calculations we have focused on two physically different examples of either ordered or disordered configurations of atoms separated by a distance $d\sim\lambda^{\mathrm{wg}}/2$, where $\lambda^{\mathrm{wg}}$ is the signal mode wavelength in the waveguide. In the case of disordered atomic chain we have simulated a particular configuration of randomly distributed atoms. The upper plot corresponds to the scattering of the signal mode, not assisted by the control field, scanned near the point of atomic resonance $\omega_0$. For the ordered configuration, when $d=\lambda^{\mathrm{wg}}/2$ precisely, there is a clear signature of cooperative enhancement of the scattering process in the backward direction, and the spectra of transmission, reflection and losses have a single resonance structure. On the contrary, for the disordered distribution the reflection is negligible and the spectra of transmission and losses have a complicated profile, which depends on the atomic configuration. However because the number of atoms is relatively small the effect of cooperativity in the entire energy balance is weak and in either case the light mostly emerges the system via incoherent scattering channel.

The lower plot of Fig.~\ref{fig4} corresponds to the scattering on an artificial resonance structure created by the control field. The calculations are presented for the control mode of frequency $\omega_c$ having the Rabi frequency $\Omega_R=2\gamma$ and tuned at the point $\Delta_c=\omega_c-\omega_0=-4\gamma$. For a sake of convenience the spectra are scaled to higher resolution and then demonstrate similar behavior as for the scattering near the original undisturbed atomic resonance shown in the upper plot. This reflects physical equivalence in manifestation of the collective effects for the scattering process from the fundamental (matter state) and artificial (driven by a control field) AT-resonance structures.

\begin{figure}[tp]
{$\scalebox{0.5}{\includegraphics*{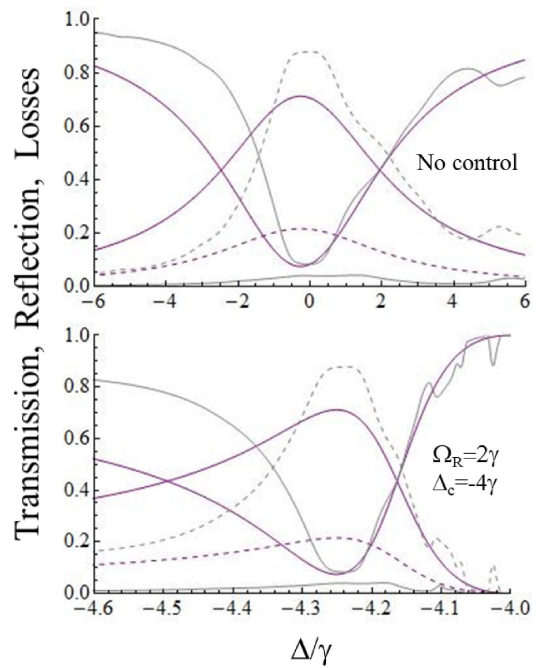}}$}
\caption{(color online) Same as in Fig.~\ref{fig4} but for an atomic chain consisting of hundred atoms.}
\label{fig5}%
\end{figure}

In Fig.~\ref{fig5} we show how the cooperative scattering in the backward direction is modified for an atomic chain consisting of hundred atoms. For a disordered configuration the backscattering and light trapping are negligible such that most of the signal light emerges the system via incoherent scattering channel and is lost. But the ordered configuration demonstrates quite strong cooperative enhancement with reduced losses, which near the resonance are even weaker than for a smaller target consisting of ten atoms, see Fig.~\ref{fig4}. Such a non-trivial spectral behavior and light trapping within the waveguide is not so contra-intuitive result and can be naturally linked with phenomena of photonic crystal and Bragg diffraction in one-dimensional and periodically structured atomic lattices. The strong coherent backscattering from ordered atomic arrays have been recently observed in experiments \cite{CGCGSKL2016,SBKISMPA2016}. The presented numerical simulations show that similar strong cooperative enhancement of the backscattering could be also observed from the artificial AT-resonance structure created by the control field.

\begin{figure}[tp]
{$\scalebox{0.5}{\includegraphics*{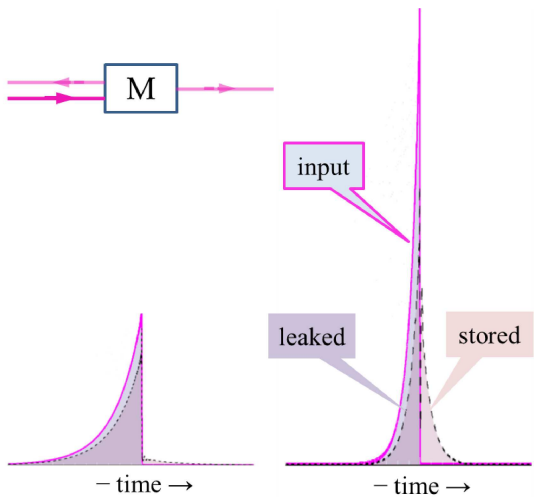}}$}
\caption{(color online) Time delay of the signal pulse, coherently scattered by the system of ten atoms (left) and hundred atoms (right). The input pulse (magenta curves) is taken as time reversed replica of the AT resonance decay profile, shown in Figs.~\ref{fig4} and \ref{fig5}. The inset shows the associated memory scheme. The outgoing pulse, with including its transmitted and reflected parts, is indicated by the dashed bounding curve. The delayed/stored fragments, responding after the back-front of the input pulse gets in the sample, are equally emitted in both the forward and backward directions. In these examples the storage efficiency is about 8\% (ten atoms) and 38\% (hundred atoms).}
\label{fig6}%
\end{figure}

The strong frequency dispersion of the signal mode can lead to delay of a signal pulse and to light storage phenomena. In Fig.~\ref{fig6} we show the memory effect for a signal pulse prepared as a time reversed replica of the AT resonance decay profile. The chosen optimally shaped signal pulse should give us a critical benchmark of the maximal memory efficiency and show entire potential for the quantum interface protocols based on a nanofiber architecture.\footnote{Strictly speaking we demonstrate here the light delay effect. But by turning off the control field at arrival time of the signal pulse and by turning it on again after delay, limited by a spin decoherence time, we would reproduce the same dynamics for the retrieved pulse excepting short and fast oscillations associated with the transient processes.} In this figure we compare the delay effect of the pulse scattered from the ordered atomic configurations consisting of ten and hundred atoms. For larger number of atoms the  cooperative interaction with the atomic chain is stronger, which makes the AT resonance broader, so the optimal signal pulse is taken shorter in time.

There are the following important properties of the process to point out. Firstly, with enlarging number of the active atoms the atomic chain tends to work as light reflector. As confirmed by our numerical simulations, in the case of hundred atoms those part of the outgoing pulse, which overlaps the input pulse and leaked from the memory protocol, is mainly scattered in the backward direction, such that for this part of the signal pulse the atomic sample indeed works as light reflector. However, that is not so for the delayed part of the light pulse, which is supposed to be stored in atomic memory. For either configurations consisting of ten, hundred or any atoms the delayed/stored part of the signal pulse is equally emitted in both directions. Secondly for the considered memory protocol the efficiency can never be perfect and in our examples it is about 8\% for ten atoms and 38\% for hundred atoms. For the infinite number of atoms it would approach to 50\% only. Below we clarify this point.

The specific interaction channels differently contribute to the construction of the outgoing pulse. The coherent Raman coupling with the control field, expressed by diagram (\ref{2.13}), provides the main cooperative response of the atomic polarization on the signal field. Nevertheless in the leaked part of the pulse there is also presence of the spontaneous Raman scattering out of the main channel within a few percents of the magnitude for both the considered atomic configurations. But the delayed/stored part of the pulse is constructed only by the cooperatively enhanced Raman emission from the repopulated (signal) atoms on the $\sigma_{-}$ optical transition, see Fig.~\ref{fig2}.

\subsection{The symmetric geometry}

\begin{figure}[tp]
{$\scalebox{0.6}{\includegraphics*{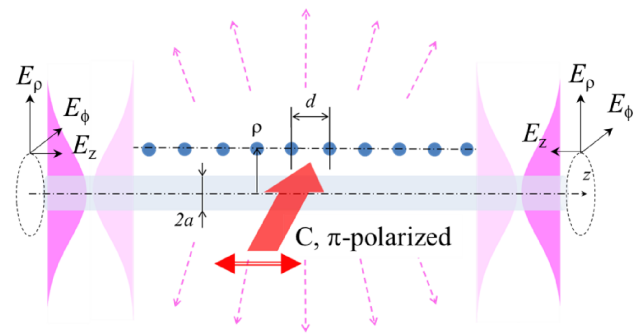}}$}
\caption{(color online). The geometry of symmetric light scattering from an array of atoms confined with a nanofiber waveguide. Unlike Fig.~\ref{fig3} here the signal light is superposed in two waveguide modes, running in opposite directions.}
\label{fig7}%
\end{figure}

\noindent Let us change the scattering geometry to the configuration when two waveguide modes, running in opposite direction, symmetrically contributes to the scattering process, see Fig.~\ref{fig7}. The signal pulse can be split and superposed in two counter-propagating fragments with aid of a ring-type interferometer based on the Sagnag design. For such a symmetric interface architecture there are two equal input channels so the target sample operates as a two-ports lossy beamsplitter for the incident light, and the scheme generalizes the conventional geometry of one-dimensional scattering, discussed in the preceding section.

The symmetric scattering process can be described by two transmission probabilities, constructed as a squared amplitudes of the signal mode passed each output channel
\begin{eqnarray}
\mathscr{T}_{\rightarrow}&=&\mathscr{T}_{\rightarrow}(\omega)=\sum_{i',k'>0}\frac{1}{2}\left|S_{i'i}+\mathrm{e}^{i\vartheta}S_{i'-i}\right|^2%
\nonumber\\%
\mathscr{T}_{\leftarrow}&=&\mathscr{T}_{\leftarrow}(\omega)=\sum_{i',k'<0}\frac{1}{2}\left|S_{i'i}+\mathrm{e}^{i\vartheta}S_{i'-i}\right|^2%
\nonumber\\%
\mathscr{L}&=&\mathscr{L}(\omega)=1-\mathscr{T}_{\rightarrow}(\omega)-\mathscr{T}_{\leftarrow}(\omega),%
\label{3.2}%
\end{eqnarray}
where by "$-i$" we have simply denoted the state, which is obtained from "$i$"-state by substituting $k\to -k$. The phase $\vartheta$ expresses the arbitrary relative phase existing between two counter-propagating incoming  fragments of the signal pulse. It is a straightforward option of the ring interferometer to further compile both the outgoing fragments with $k'>0$ and $k'<0$ in one pulse.

\begin{figure}[tp]
{$\scalebox{0.5}{\includegraphics*{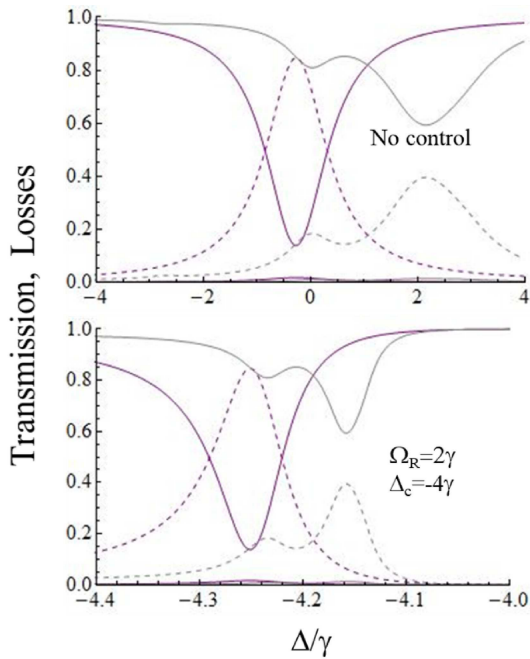}}$}
\caption{(color online) Spectra of transmission $\mathscr{T}=\mathscr{T}(\omega)\equiv\mathscr{T}_{\rightarrow}(\omega)+\mathscr{T}_{\leftarrow}(\omega)$ and losses $\mathscr{L}=\mathscr{L}(\omega)$ of light scattered from a symmetrically irradiated atomic chain consisting of ten atoms, which were calculated without (upper plot) and in the pesence (lower plot) of the control field. Other parameters and curve specifications are the same as in Fig.~\ref{fig4}.}
\label{fig8}%
\end{figure}%

The specifics of scanning the atomic sample in symmetric geometry is clear seen for a periodically ordered atomic configuration. In this case the driving $\sigma_{-}$-polarized component of the incident wave is given by superposition of two counter-propagating running waves (see Fig.~\ref{fig1}) with equal amplitudes and creates a standing wave.  By varying the phase $\vartheta$ the atoms could occupy either the nodes or crests of such standing wave.  In the former case the system would be completely transparent with $\mathscr{T}_{\rightarrow}=\mathscr{T}_{\leftarrow}=1/2$ such that the presence of atoms would be invisible for the signal light. But in the latter case, with placing atoms in the crests, the interaction would be maximally enforced, and below we present our numerical simulations just for such an advanced design.

In Fig.~\ref{fig8} we show the spectra of total transmission $\mathscr{T}(\omega)\equiv\mathscr{T}_{\rightarrow}(\omega)+\mathscr{T}_{\leftarrow}(\omega)$ and losses $\mathscr{L}(\omega)$ for the scattering from the symmetrically irradiated atomic sample, consisting of ten atoms. The main features of the process are the same as they are for the conventional scattering process, which spectra are shown in Fig.~\ref{fig4}. As before the spectral profiles of the undisturbed fundamental atomic resonance (upper plot) and of the artificial AT resonance (lower plot) demonstrate certain similarity of their spectral shapes. For disordered configuration the atoms scatter the light mostly via incoherent channel, so that in Fig.~\ref{fig8} we obtain the balance between transmission and losses approximately at the same level as in Fig.~\ref{fig4}. It cannot be point by point coincidence between the graphs since the conventional scattering process, when light is arriving from different directions, is not symmetric in the disordered case. On the contrary, the ordered configuration reveals important difference between the respective dependencies in Fig.~\ref{fig8} and Fig.~\ref{fig4}, so for symmetric scanning the losses are evidently higher and transmission is weaker. That seems as an expectable consequence of the above pointed enforcing of the interaction between the light and atoms for this case.

\begin{figure}[tp]
{$\scalebox{0.5}{\includegraphics*{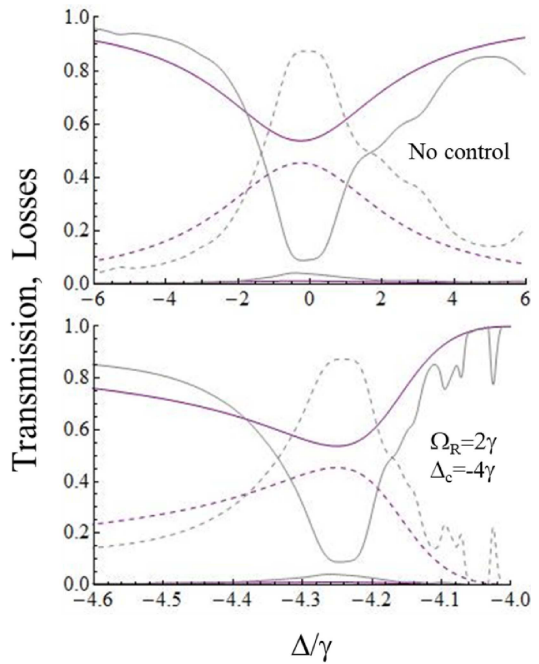}}$}
\caption{(color online) Same as in Fig.~\ref{fig8} but for hundred atoms.}
\label{fig9}%
\end{figure}%

In Fig.~\ref{fig9} we show how the spectra of transmission and losses are modified for an atomic sample consisting of hundred atoms. For disordered configuration we obtain much stronger scattering and losses via incoherent channel. But the ordered atomic chain demonstrates different behavior showing that the atomic sample becomes more transparent than for the chain consisting of less atoms. From the first sight this surprising result may seem as contra-intuitive and contradicting the above arguments. But it is clarified once we pay attention that with enlarging number of atoms the scattering process turns on to the cooperative dynamics with enhancement of the Purcell effect such that most of the light is re-emitted back to the guided mode as result of cooperative either Rayleigh (upper plot in Fig.~\ref{fig9}) or Raman (lower plot in Fig.~\ref{fig9}) scattering processes. If we compare the dependencies of Fig.~\ref{fig9} with the similar dependencies of Fig.~\ref{fig5} we can conclude that for the symmetrically organized scattering process the atomic chain, consisting of a sufficiently large number of atoms, should asymptotically approach the properties of lossless beamsplitter capable only to redirect the light beams within one-dimensional channel.

\begin{figure}[tp]
{$\scalebox{0.5}{\includegraphics*{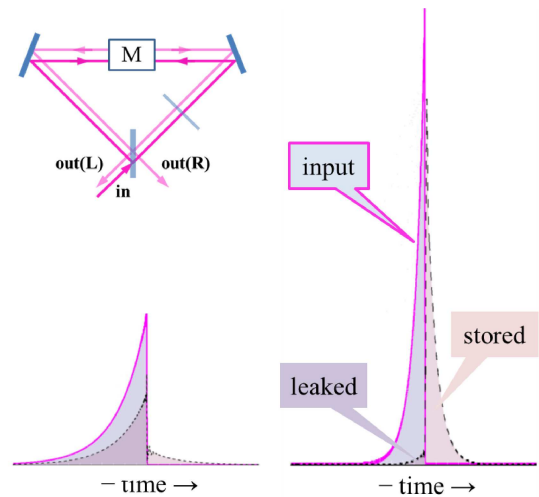}}$}
\caption{(color online) Same as in Fig.~\ref{fig6} but for the symmetric scattering. The inset shows possible observation scheme based on a ring-type interferometer with a half-wave plate placed in its right-hand arm.  The signal pulse, retrieved from the memory unit $M$, emerges from both the beamsplitter ports $\mathrm{out}(L)$ and $\mathrm{out}(R)$ and is further detected. In these examples the storage efficiency is within 16\% (ten atoms) and 76\% (hundred atoms).}
\label{fig10}%
\end{figure}

As a consequence, the symmetric geometry should lead to more effective memory effect since for a periodically structured atomic chain the pulse retrieval, provided by the cooperative Raman emission, would be intrinsically symmetric process. In Fig.~\ref{fig10} we show how the split signal pulse impinging the atomic sample from the opposite sides could be delayed and stored in the spin subsystem. For the considered examples the efficiency of the memory is about 16\% for ten atoms and 76\% for hundred atoms so it is exactly twice more than for the conventional scattering design when the pulse would arrive from only one side, see Fig.~\ref{fig6}. The inset in Fig.~\ref{fig10} suggests possible scheme of a ring-type interferometer adjusted for the observation of the process. It needs to place a half-wave plate in the right-hand interferometer arm to change helicity of the wave before its conversion into the waveguide mode.  Although in such a memory scheme there is an option that the retrieved pulse would outcome the system from both the interferometer ports, our calculations confirms that the retrieved photon emerges the same port as it has arrived.

With enlarging number of atoms up to infinity the efficiency would approach 100\%. Now we can see that for the conventional geometry the maximal efficiency is limited by 50\% just because the incoming pulse can be optimized as a time reversed replica for the cooperatively emitted light only in one direction i. e. for only half energy of the potentially stored light. As a clear visualization of the time reversal symmetry for a sample with large number of atoms the incoming and stored pulses, considered together, construct the symmetric time profile, see the right image in Fig.~\ref{fig10}.

Here we have demonstrated the memory effect by considering direct Raman conversion of the signal pulse onto the spin excitation. But as commented in \cite{Gorshkov2007}, for the simplest $\Lambda$-configured atoms the memory protocol, developed near the point of electromagnetically induced transparency (EIT), should have the same efficiency. In the next section we show that in the limit of strong cooperativity for a one-dimensional system of the tripod atoms there is indeed similarity in description of either absorption (AT) or transparency (EIT) resonances, so both the light storage protocols can be equally applicable.

\subsection{The two-channel model}

\noindent Imagine that the Purcell enhancement of the radiation emission is as strong as the atoms mostly communicate via the guided mode. In such a highly effective interface configuration the above results can be approximated by a simplified model with keeping only the main scattering channels. Accordingly to the basic statements of the quantum scattering theory, see \cite{MottMassey85}, the two channel process can be described by the scattering matrix given by the product
\begin{equation}
S=\left(\begin{array}{cc} \cos\alpha & -\sin\alpha \\ \sin\alpha & \cos\alpha \end{array}\right)\!%
\left(\begin{array}{cc} e^{2i\delta_1} & 0 \\ 0 & e^{2i\delta_2} \end{array}\right)\!%
\left(\begin{array}{cc} \cos\alpha & \sin\alpha \\ -\sin\alpha & \cos\alpha \end{array}\right)\!,%
\label{3.3}
\end{equation}
which makes general factorization of a symmetric unitary matrix in two dimensional subspace. The incoming and outgoing channels $1$ and $2$ should be associated here with two counter-propagating waves either incident on or scattered from the atomic target.

As follows from the symmetry of the considered system there are two combinations of the waves incident from opposite directions, which would independently interact with the target sample. These are either symmetric or antisymmetric superpositions of the counter-propagating incident waves. In respect to the frame, centered within the sample, these two wave configurations are shifted by $\lambda^{\mathrm{wg}}/4$ scale so the atoms are distributed either in the crests or in the nodes of the interfering waves. In the latter case there is no interaction between light and atoms such that the ordered atomic array becomes completely transparent. But in the former case the system has an ability of strong cooperative emission into the outgoing modes. Thus the $S$-matrix (\ref{3.2}) could be parameterized by $\alpha=\pi/4$ and $\delta_2=0$ (no scattering). Near an isolated point of the resonance scattering the exponent with the phase shift $\delta_1$ can be expressed as
\begin{equation}
e^{2i\delta_1}=\pm\frac{\delta\omega-\frac{i}{2}\Gamma_{C}}{\delta\omega+\frac{i}{2}\Gamma_{C}}\, ,%
\label{3.4}%
\end{equation}
where we have denoted $\delta\omega=\omega-\omega_{\ast}$ and $\omega_{\ast}$ is the point of resonance, which can be associated with either absorption AT-resonance or transparency EIT-resonance. The positive and negative signs are respectively related with the tuning at the points of these resonances and for a sake of simplicity we have parameterized both the resonances by the same cooperative width $\Gamma_{C}$. In our previous discussion $\Gamma_{C}$ was given by an effective rate of the cooperative Raman emission into the outgoing channel.

For positive sign and for tuning the signal mode near the point of absorbtion resonance $\delta\omega\sim 0$ the system transfers to a perfect phase inverting reflector, i.e. matrix (\ref{3.2}) becomes cross-diagonal with off-diagonal matrix elements equal to $-1$. In opposite case, with keeping negative sign in (\ref{3.2}), near the point $\delta\omega\sim 0$ the system becomes transparent such that the $S$-matrix transforms to identity matrix. As we can see, for explanation of the light storage effect there is no difference between these two types of resonances if both are parameterized by the same line width $\Gamma_{C}$. And it is quite intriguing that such an ideal scenario is reproducible by our precise numerical simulations.

The representation of $S$-matrix in the form of isolated resonance (\ref{3.3}) and (\ref{3.4}) is as better valid as the resonance can be reliably approximated by a Lorentzian shape profile. In the presented calculations both the polariton modes, associated with the pole structure of the resolvent operator either at the point of atomic resonance or near the detuning of the control field, are fairly resolved and such an approximation seems faithful. The scaling of the effective resonance bandwidth is given by $\Gamma_C\sim \gamma^{\mathrm{wg}}N$ where $\gamma^{\mathrm{wg}}$ is a dimensional constant, which could be formally interpreted as emission rate into the waveguide per atom and confirm our expectations preceded in the introduction. Nevertheless we can point out intrinsically quantum nature of the process when the polariton modes are expressed by an entangled non-separable Dicke-type microscopic quantum states. The polariton dynamics cannot be visualized as a subsequent excitation transfer along the atomic chain. The periodically ordered atoms exist in equivalent physical conditions such that they simultaneously interact each other within any pair with the same coupling strength.

That would be not the case for a disordered atomic chain where incoherent losses would dominate in the scattering process. Any signatures of the localization phenomenon, possibly associated with the disorder induced light trapping inside the atomic chain, are strongly suppressed because of the losses. The light propagates through a disordered one-dimensional atomic chain similarly to light transport through a dilute atomic gas and the process implies the standard self-consistent description in terms of macroscopic Maxwell approach. The basic features of the Raman process, including the EIT effect, can be correctly introduced in terms of nonlinear susceptibility of the atomic sample, Beer-Lambert-Bouguer law, slow light etc..

\begin{figure}[tp]
{$\scalebox{0.45}{\includegraphics*{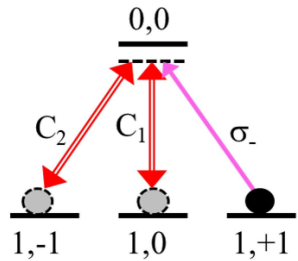}}$}
\caption{(color online). Transition diagram showing how two control fields $C_1$ and $C_2$ having different polarizations can be used for storage of two signal photons in a multiatomic chain.}
\label{fig11}%
\end{figure}

Since in the typical experimental conditions a number of atoms, contributing to the process, can be sufficiently large the nanofiber systems show good potential to be extended towards various other more sophisticated interface schemes. As example in Fig.~\ref{fig11} it is shown how two photons could be subsequently stored in an atomic array with make use of two control fields of different polarizations. The $\sigma_{+}$ polarized control mode is assumed as a plane wave directed at small angle to the waveguide ($z$-axis). The retrieval of the photons can be done either independently or simultaneously if we would aim for preparation of a two photon state on demand. Furthermore it seems quite straightforward that with manipulation by mode polarizations and with involving the different segments of the entire atomic array, normally consisting of thousands atoms, one can adjust the proposed design for the quantum information processing of many qubits.

\section{Conclusion}

\noindent In this paper we have examined the scheme of quantum interface based on coherent interaction of a single photon pulse with a one-dimensional atomic system. The key point of our analysis was in realistic modeling of the interaction process not restricted by simplifying approximations of two-level atoms, hopping transfer of an optical excitation, empirical description of the scattering sequence from atomic chain etc. As a clarifying example, attainable for the existing experimental capabilities, we have considered an array of the tripod-type atoms confined with a nanoscale waveguide and interacting with the signal light propagating in its fundamental HE${}_{11}$-mode.

It has been obtained that cooperative phenomena strongly affect the interaction process. The array of atoms periodically ordered along the waveguide and separated by a distance of half wavelength has ability of cooperative enhancement of the radiation emission into the guided mode. For a sufficiently but not extremely large number of the trapped atoms (taken about hundred in our numerical simulations) the system tends to be adjusted as an effective beamsplitter redirecting the light within the waveguide. In a conventional scattering configuration the signal light it mostly reflected at the resonance point with strong dispersion for the polariton mode created in the entire system. This effect can be observed not only near the fundamental matter state atomic resonance but also at the artificial Autler-Townes resonance, structured and manipulated by the control field.

From the point of view of quantum interface we have verified that it would be optimal that the signal light was symmetrically incident from both the sides of the atomic sample. Owing to periodic symmetry in the atomic distribution the counter-propagating polariton modes would equally responde the optical excitation of the sample in this case. Our supporting numerical simulations have confirmed the strong scattering, strong dispersion and existence of the memory effect. For most effective quantum state interchange between the light and atoms the temporal profile of the signal pulse should be shaped as a time reversed copy of the decay profile for the Autler-Townes resonance. The considered one-dimensional scheme of quantum memory is attainable for experimental verification and has a certain potential towards design of a scalable  multiqubit quantum interface, quantum registers and logic operations.

\section*{Acknowledgements}
\noindent This work was supported by the Russian Foundation for Basic Research under Grants \# 18-02-00265-A and \#19-52-15001-CNRS-a, by the Russian Science Foundation under Grant \# 18-72-10039, and by the Foundation for the Advancement of Theoretical Physics and Mathematics "BASIS" under Grant \# 18-1-1-48-1.
\appendix

\section{The Green's function $D^{(\mathrm{ext})}$}\label{Appendix_A}

\noindent As shown in \cite{PLGPCLK2018} the contribution to the Green's function $D^{(\mathrm{ext})}$, physically associated with the radiation emitted by a point-like dipole source into external modes, can be approximated by the following correction to the vacuum Green's function
\begin{eqnarray}
\lefteqn{\hspace{-0.5cm}D^{(\mathrm{ext})}_{\mu\nu}(\mathbf{r},\mathbf{r}';\omega)\approx\tilde{D}^{(0)}_{\mu\nu}(\mathbf{r},\mathbf{r}';\omega)\equiv D^{(0)}_{\mu\nu}(\mathbf{r}-\mathbf{r}';\omega)}%
\nonumber\\%
&&\hspace{-1cm}-\sum_{s}E^{(s)}_{\mu}(\mathbf{r})\!\int\!\!d^3r''\,D^{(s)\ast}_{\alpha}(\mathbf{r}'')D^{(0)}_{\alpha\nu}(\mathbf{r}''-\mathbf{r}';\omega).%
\label{a.1}
\end{eqnarray}
Here $D^{(0)}_{\mu\nu}(\mathbf{r},\mathbf{r}';\omega)$ is the fundamental solution of the wave equation in free space, and $E^{(s)}_{\mu}(\mathbf{r})$, $D^{(s)}_{\alpha}(\mathbf{r}'')$ are the vector components respectively for the electric and displacement fields of the HE${}_{11}$-mode. This approximation is as better valid as the dipole source is farther separated from the fiber surface. We will further assume that the dipole has a position within a tail of the evanescent field where the fundamental waveguide mode can be faithfully approximated by a paraxial Gaussian mode propagating in free space.

The electric field Green's function in free space is defined as solution of the microscopic Maxwell equation
\begin{eqnarray}
\lefteqn{\triangle D^{(0)}_{\mu\nu}(\mathbf{r},\mathbf{r}';\omega) -\frac{\partial^2}{\partial x_{\mu}\partial x_{\alpha}}D^{(0)}_{\alpha\nu}(\mathbf{r},\mathbf{r}';\omega)}
\nonumber\\%
&+&4\pi\frac{\omega^2}{c^2}D^{(0)}_{\mu\nu}(\mathbf{r},\mathbf{r}';\omega)=4\pi\hbar\,\frac{\omega^2}{c^2}\delta_{\mu\nu}\delta(\mathbf{r}-\mathbf{r}'),%
\nonumber\\%
\label{a.2}
\end{eqnarray}
which can be constructed with Fourier transform
\begin{eqnarray}
D^{(0)}_{\mu\nu}(\mathbf{R};\omega)&=&\int\!\frac{d^3\kappa}{(2\pi)^3}\,\mathrm{e}^{i\boldsymbol{\kappa}\cdot\mathbf{R}}
\nonumber\\%
&&\times\frac{4\pi\hbar\omega^2}{\omega^2-\omega_{\kappa}^{2}+i0}\left[\delta_{\mu\nu}-c^2\frac{\kappa_{\mu}\kappa_{\nu}}{\omega^2}\right],%
\label{a.3}
\end{eqnarray}
where $\mathbf{R}=\mathbf{r}-\mathbf{r}'$. The internal wave vector argument can be decomposed as $\boldsymbol{\kappa}=(\boldsymbol{\kappa}_{\perp}=\mathbf{q},\kappa_z=k)$ and for the mode frequency one has $\omega_{\kappa}^2=c^2k^2+c^2\mathbf{q}^2$.

In (\ref{a.1}) the displacement field $\mathbf{D}^{(s)}(\mathbf{r})$ is considered in Cartesian basis
\begin{equation}
\mathbf{D}^{(s)}(\mathbf{r})=D^{(s)}_{\alpha}(\mathbf{r})\,\mathbf{e}_{\alpha}=D^{(s)}_{\alpha}(\rho,\phi)\,\frac{1}{\sqrt{{\cal L}}}\mathrm{e}^{ik z}\,\mathbf{e}_{\alpha}%
\label{a.4}
\end{equation}
with $\alpha=x,y,z$ and the integral in the second line of (\ref{a.1}) can be transformed as
\begin{eqnarray}
\lefteqn{\int\!\!d^3r''\,D^{(s)\ast}_{\alpha}(\mathbf{r}'')D^{(0)}_{\alpha\nu}(\mathbf{r}''-\mathbf{r}';\omega)}%
\nonumber\\%
&&=\frac{1}{\sqrt{{\cal L}}}\,\mathrm{e}^{-ik z'}\!\int\!d^2\!\rho''\,D^{(s)\ast}_{\alpha}(\rho'',\phi'')\int\!\frac{d^2q}{(2\pi)^2}\,%
\mathrm{e}^{i\mathbf{q}\cdot(\boldsymbol{\rho}''-\boldsymbol{\rho}')}
\nonumber\\%
&&\times\frac{4\pi\hbar\omega^2}{\omega^2-c^2k^2-c^2\mathbf{q}^2+i0}\left[\delta_{\alpha\nu}-c^2\frac{\kappa_{\alpha}\kappa_{\nu}}{\omega^2}\right]%
\nonumber\\%
&&=\frac{1}{\sqrt{{\cal L}}}\,\mathrm{e}^{-ik z'}\int\!\frac{d^2q}{(2\pi)^2}\mathrm{e}^{-i\mathbf{q}\cdot{\mathbf{\rho}'}}\,D^{(s)\ast}_{\alpha}(\mathbf{q})%
\nonumber\\%
&&\times\frac{4\pi\hbar\omega^2}{\omega^2-c^2k^2-c^2\mathbf{q}^2+i0}\left[\delta_{\alpha\nu}-c^2\frac{\kappa_{\alpha}\kappa_{\nu}}{\omega^2}\right],
\label{a.5}
\end{eqnarray}
where we have used cylindrical coordinates, with $\mathbf{r}=(\boldsymbol{\rho},z)=(\rho,\phi,z)$ and $d^2\rho=\rho d\rho d\phi$, and defined the Fourier components of the displacement field in respect to its transverse spatial coordinates
\begin{equation}
D^{(s)}_{\alpha}(\mathbf{q})=\int\!d^2\!\rho\;\mathrm{e}^{-i\mathbf{q}\cdot\boldsymbol{\rho}}\,D^{(s)}_{\alpha}(\rho,\phi),%
\label{a.6}
\end{equation}
which contributes in (\ref{a.5}) in its complex conjugated form.

The basic expansion for the HE${}_{11}$-mode (\ref{2.12}), presented for the vector of displacement field, can be written as~\footnote{For a sake of convenience we have multiplied the complex components (\ref{2.12}) (main text) by an extra phase factors "$-i$". These auxiliary factors compensate each other and can be canceled out in the final expressions (\ref{a.13})-(\ref{a.15})}
\begin{eqnarray}
\mathbf{D}^{(s)}(\boldsymbol{\rho})&=&D_{\perp}(\rho)\,\mathbf{e}_s+\mathbf{D}^{(s)}_{2\phi}(\rho,\phi)+\mathbf{D}^{(s)}_{z}(\rho,\phi)%
\nonumber\\%
\mathbf{D}^{(s)}(\mathbf{q})&=&D_{\perp}(q)\,\mathbf{e}_s+\mathbf{D}^{(s)}_{2\phi}(\mathbf{q})+\mathbf{D}^{(s)}_{z}(\mathbf{q}).%
\label{a.7}
\end{eqnarray}
For a sub-wavelength waveguide the first term dominates and implies description of paraxial optics so $\mathbf{e}_s$ is the unit complex polarization vector of the mode in its paraxial limit. By this we mean that the mode can be faithfully approximated by the the paraxial solution of the wave equation in free space. We will mainly track in the mode index $s=\sigma,k$ its azimuthal number $\sigma=\pm 1$, which in this limit corresponds to either right-handed or left-handed circular polarizations. In accordance with (\ref{2.10})-(\ref{2.12}) the first term in (\ref{a.7}) is given by
\begin{eqnarray}
D_{\perp}(\rho)&=&\epsilon(\rho)\;\frac{-iE_{\rho}(\rho)-E_{\phi}(\rho)}{2\sqrt{\pi}}%
\nonumber\\%
D_{\perp}(q)&=&2\pi\!\int_0^\infty\!\!\rho d\rho\, J_0(q\rho)\,D_{\perp}(\rho),%
\label{a.8}
\end{eqnarray}
where $\epsilon(\rho)$ is dielectric permittivity of the entire medium (dielectric fiber and free space).

The second term in (\ref{a.7}) is a specific waveguide contribution, which depends on azimuthal angle $\phi$ and vanishes in the paraxial limit. It is given by
\begin{eqnarray}
\left.\mathbf{D}^{(s)}_{2\phi}(\rho,\phi)\right|_{\sigma=\pm 1}&=&\mathbf{e}_{\bar{s}}D'(\rho)\mathrm{e}^{\pm 2i\phi}%
\nonumber\\%
D'(\rho)&=&\epsilon(\rho)\;\frac{-iE_{\rho}(\rho)+E_{\phi}(\rho)}{2\sqrt{\pi}}%
\nonumber\\%
\left.\mathbf{D}^{(s)}_{2\phi}(\mathbf{q})\right|_{\sigma=\pm 1}&=&\mathbf{e}_{\bar{s}}\,2\pi (\pm i)^2\!\int_0^\infty\!\!\rho d\rho\, J_{\pm 2}(q\rho)\,D'(\rho)%
\nonumber\\%
&\equiv&-\mathbf{e}_{\bar{s}}D'(q),%
\label{a.9}
\end{eqnarray}
where $\mathbf{e}_{\bar{s}}$ is the unit transverse complex vector orthogonal to $\mathbf{e}_s$ ($\mathbf{e}_s^{\ast}\cdot\mathbf{e}_{\bar{s}}=0$) and $z$-axis.

The last term in (\ref{a.7}) defines the field longitudinal component, also vanishing in the paraxial limit, and it is given by
\begin{eqnarray}
\left.\mathbf{D}^{(s)}_{z}(\rho,\phi)\right|_{\sigma=\pm 1}&=&\mathbf{e}_{z}\,(-i)D''(\rho)\mathrm{e}^{\pm i\phi}%
\nonumber\\%
D''(\rho)&=&\epsilon(\rho)\,\frac{E_z(\rho)}{\sqrt{2\pi}}%
\nonumber\\%
\left.\mathbf{D}^{(s)}_{z}(\mathbf{q})\right|_{\sigma=\pm 1}&=&\mathbf{e}_{z}\, 2\pi(\pm i) \!\int_0^\infty\!\!\rho d\rho\, J_{\pm 1}(q\rho)\,(-i)D''(\rho)%
\nonumber\\%
&\equiv& \mathbf{e}_{z}D''(q),%
\label{a.10}
\end{eqnarray}
where $\mathbf{e}_{z}$ is unit vector along the z-axis. The last lines of Eqs.~(\ref{a.9}) and (\ref{a.10}) are constituted by the property of the Bessel functions: $J_{-m}(x)=(-)^mJ_m(x)$. All the functions $D_{\perp}$, $D'$ and $D''$ are set as real functions.

The integral (\ref{a.5}), as being substituted to (\ref{a.1}), has to be evaluated for near resonance waveguide modes i.e. for those longitudinal wave numbers $k$, which fulfil the condition: $\omega_s\sim\omega\sim\omega_0$. It is a consequence of that physically just these waveguide modes are mainly responding on radiation resonantly scattered by the atoms. Since the phase and group velocities of the waveguide mode are always less than speed of light in vacuum we get $\omega^2-c^2k^2\sim\omega_s^2-c^2k^2<0$ and in these conditions the denominator in (\ref{a.5}) is off-resonant and always negative for any $\mathbf{q}^2$ such that the integral is uniformly convergent inside the spectral domain where function $\mathbf{D}^{(s)\ast}(\mathbf{q})$ is meaningful. Furthermore the denominator in the integrand additionally filters the transverse modes contributing to integral (\ref{a.5}) at the scale $q^2\sim (c^2k^2-\omega^2)/c^2$. The filtering function selects a spectral area near the frame origin in the reciprocal space i.e. at $|\mathbf{q}|=q\to 0$. Under our assumptions the scaling of $q$ is within deviation of $\omega\sim\omega_s\sim\omega_0$ from $c\,k\equiv\omega_k^{\mathrm{free}}$. Typically the difference $\omega_k^{\mathrm{free}}-\omega_s\ll\omega_s$ is sufficiently small but at the same time much larger than the width of atomic resonance $\omega_k^{\mathrm{free}}-\omega_s\gg\gamma$.

The terms $\mathbf{D}^{(s)\ast}_{2\phi}(\mathbf{q})$  and $\mathbf{D}^{(s)\ast}_{z}(\mathbf{q})$ approach zero at the origin point, see Eqs.~(\ref{a.9}), (\ref{a.10}), so their contribution to the convolution-type integral (\ref{a.5}) is suppressed by the filtering function, and the above estimates let us neglect them. Physically that means that the convolution of the waveguide mode with the vacuum Green's function selects those contribution to the mode which can be approximated by the transverse Gaussian waves propagating in free space. With leaving in the integrand of (\ref{a.5}) only the contribution surviving the limit $q\to 0$ (i.e. with omitting contributions of $\mathbf{D}^{(s)\ast}_{2\phi}(\mathbf{q})$ and $\mathbf{D}^{(s)\ast}_{z}(\mathbf{q})$ as well as the second term in the square brackets of (\ref{a.5})) we arrive at the following estimate of our basic integral
\begin{eqnarray}
\lefteqn{\int\!\!d^3r''\,D^{(s)\ast}_{\alpha}(\mathbf{r}'')D^{(0)}_{\alpha\nu}(\mathbf{r}''-\mathbf{r}';\omega)}%
\nonumber\\%
\nonumber\\%
&&\approx\frac{1}{\sqrt{{\cal L}}}\,\left(\mathbf{e}_{s}^{\ast}\right)_{\nu}\mathrm{e}^{-ik z'}\int_0^\infty\!\frac{qdq}{2\pi}\,J_0(q\rho')\,D_{\perp}(q)%
\nonumber\\%
&&\times\frac{4\pi\hbar\omega^2}{\omega^2-c^2k^2-c^2q^2+i0}.%
\label{a.11}
\end{eqnarray}
We have justified this approximation for the wave numbers $k$ obeying the condition $\omega_s\sim\omega\sim\omega_0$ where we have relevant balance between the spectral parameters. Nevertheless we formally extend it over all the $k$'s and evaluate the integral (\ref{a.1}) formally in infinite limits. The result is correct inside the frequency domain where the respective waveguide modes $\omega_s\sim\omega$ mainly exist in evanescent field outside the fiber and can be reproducible by paraxial Gaussian modes propagating in free space. It is incorrect for high frequencies, where the modes are concentrated inside the fiber with minimal contribution of the evanescent field. But in such a situation the approximation (\ref{a.1}) would be insufficient for itself.

After substitution (\ref{a.11}) into (\ref{a.1}) the Green's function $D^{(\mathrm{ext})}$ is given by
\begin{eqnarray}
\lefteqn{\hspace{-0.5cm} D^{(\mathrm{ext})}_{\mu\nu}(\mathbf{r},\mathbf{r}';\omega)\approx D^{(0)}_{\mu\nu}(\mathbf{r}-\mathbf{r}';\omega)}%
\nonumber\\%
\nonumber\\%
&&\hspace{-0.3cm}+\;E^{(s)}_{\mu}(\rho,\phi)\,\left(\mathbf{e}_{s}^{\ast}\right)_{\nu}%
\int_0^\infty\!\frac{qdq}{2\pi}\,J_0(q\rho')\,D_{\perp}(q)%
\nonumber\\%
&&\hspace{-0.3cm}\times\,\frac{2\pi i\,\hbar\omega^2}{c\sqrt{\omega^2-c^2q^2}}\ \exp\left[i\sqrt{\omega^2/c^2-q^2}\,|z- z'|\right].
\label{a.12}
\end{eqnarray}
Here we have evaluated the integral over longitudinal wavenumber in assumption that the transverse profile of the field is approximately independent on $k$ in the representative area of integration including the poles $k\sim\pm\omega/c$.  In (\ref{a.12}) and below in (\ref{a.13}), (\ref{a.15}) index "$s$" enumerates only azimuthal number $\sigma$ of the mode together with specification of its propagation direction (forward for $z>z'$ and backward for $z<z'$) and assumes the sum over the repeated index in the product.  The remaining integral cannot be evaluated in a closed form, but it can be conveniently approximated by its near and far distant asymptotes.

For short separations $|z-z'|\lesssim \omega/c\overline{q^2}$, where $1/\overline{q^2}$ is inverse variance of the transverse wave number, one could ignore $q^2$ in the second line of the integrand and get
\begin{eqnarray}
\lefteqn{\left.D^{(\mathrm{ext})}_{\mu\nu}(\mathbf{r},\mathbf{r}';\omega)\right|_{|z-z'|< 2\omega/c\overline{q^2}}\sim D^{(0)}_{\mu\nu}(\mathbf{r}-\mathbf{r}';\omega)\ }%
\nonumber\\%
\nonumber\\%
&&+\ E^{(s)}_{\mu}(\rho,\phi)\left(\mathbf{e}_{s}^{\ast}\right)_{\nu}D^{\ast}_{\perp}(\rho')\;\frac{2\pi i\,\hbar\omega}{c}\ \exp\left[i\frac{\omega}{c}\,|z- z'|\right],%
\nonumber\\%
\label{a.13}
\end{eqnarray}
where we have returned the original phase convention for the mode definition in accordance with (\ref{2.12}) and made use of completeness of the Bessel functions.\footnote{After substituting (\ref{a.8}) into (\ref{a.12}): $$\int_0^\infty\!qdq\,J_0(q\rho')\,J_0(q\rho)=\frac{1}{\rho}\,\delta(\rho'-\rho),$$ see G.N. Watson, \textit{A Treatise on the Theory of Bessel Functions} (Cambridge University Press, Cambridge,U.K., 1966).} In the area outside the fiber, where the electric and displacement field coincide, this result can be rewritten in equivalent form
\begin{eqnarray}
\lefteqn{\hspace{-0.7cm}\left.D^{(\mathrm{ext})}_{\mu\nu}(\mathbf{r},\mathbf{r}';\omega)\right|_{|z-z'|< 2\omega/c\overline{q^2}}\approx D^{(0)}_{\mu\nu}(\mathbf{r}\!-\!\mathbf{r}';\omega)}%
\nonumber\\%
\nonumber\\%
&&\hspace{-0.7cm}-\ \sum_{s}\frac{4\pi\hbar\omega^2}{\omega^2-c^{2}k^{2}+i0}\,E^{(s)}_{\mu}(\mathbf{r})\,E^{(s)\ast}_{\bot\nu}(\mathbf{r}'),%
\label{a.14}
\end{eqnarray}
where in the subtracting term $\mathbf{E}^{(s)}_{\bot}(\mathbf{r}')$ denotes the leading (in paraxial limit) contributions in the right-hand side of (\ref{a.7}) when the terms depended on azimuthal angle are omitted.

The physical consistence of the obtained result was commented in appendix A of \cite{PLGPCLK2018}. The inverse variance of the transverse wave number $1/\overline{q^2}\sim w^2$ gives us an estimate for the beam waist $w$ and $\omega/c=2\pi/\lambda_0$ defines the vacuum wavelength $\lambda_0$ which differs from the waveguide wavelength $\lambda^{\mathrm{wg}}$, but in accordance with our basic assumptions $\lambda_0-\lambda^{\mathrm{wg}}\ll\lambda^{\mathrm{wg}}$. In diffraction theory the length scale $\pi w^2/\lambda_0= z_R$ is known as Rayleigh range. Thus the derived approximation for the Green's function $D^{(\mathrm{ext})}$ (\ref{a.14}) is applicable for the radiation coupling between the atoms separated by a distance within double Rayleigh range associated with a paraxial Gaussian fit of the HE${}_{11}$-mode.

In alternative limit $|z-z'|\gg \omega/c\overline{q^2}$ the oscillating exponent would reduce the integral in (\ref{a.12}) down to zero value. For separations $\rho'\lesssim c/\omega\ll \omega/c\overline{q^2}$ in the rest it has the following asymptote
\begin{eqnarray}
\lefteqn{\hspace{-1.5cm}\left.D^{(\mathrm{ext})}_{\mu\nu}(\mathbf{r},\mathbf{r}';\omega)\right|_{|z-z'|\gg \omega/c\overline{q^2}}\sim D^{(0)}_{\mu\nu}(\mathbf{r}-\mathbf{r}';\omega)}%
\nonumber\\%
\nonumber\\%
&&\hspace{-1.5cm}+\ E^{(s)}_{\mu}(\rho,\phi)\left(\mathbf{e}_{s}^{\ast}\right)_{\nu}\,\int_0^\infty\!\!2\pi\rho''d\rho''\,D_{\perp}^{\ast}(\rho'')%
\nonumber\\%
&&\times\ \frac{\hbar\omega^2}{c^2|z- z'|}\ \exp\left[i\frac{\omega}{c}\,|z- z'|\right],%
\label{a.15}
\end{eqnarray}
where the second term subtracts the dipole emission into a small solid angle overlapping the area shining by the waveguide mode.

\bibliographystyle{apsrev4-1}
\bibliography{references}

\end{document}